\documentclass[fleqn,usenatbib]{mnras}

\usepackage{newtxtext,newtxmath}

\usepackage[T1]{fontenc}

\DeclareRobustCommand{\VAN}[3]{#2}
\let\VANthebibliography\thebibliography
\def\thebibliography{\DeclareRobustCommand{\VAN}[3]{##3}\VANthebibliography}

\usepackage{graphicx}	
\usepackage{amsmath}	
\usepackage{placeins}
\usepackage{multicol}
\usepackage{xcolor}

\newcommand{\kms}{~km\,s$^{-1}$}

\title[Submillimeter methanol masers near S255 NIRS3]{{Submillimeter Class II methanol masers near the massive protostar S255IR NIRS3: evolution and excitation of the $J_1 -J_0$ A$^{-+}$ series and a new maser line at 345.919~GHz}}

\author[I. I. Zinchenko et al.]{
I. I. Zinchenko,$^{1}$\thanks{E-mail: zin@ipfran.ru}
S. V. Salii,$^{2}$
A. M. Sobolev,$^{2}$
I. A. Zaichikova,$^{1}$
S.-Y. Liu,$^{3}$
and Y.-N. Su$^{3}$
\\
$^{1}$Federal Research Center A.V. Gaponov-Grekhov Institute of Applied Physics of the Russian Academy of Sciences, \\ 46 Ul’yanov str., Nizhny Novgorod 603950, Russia\\
$^{2}$Astronomical Observatory, Ural Federal University, 51, Lenin ave.,  Ekaterinburg 620000, Russia\\
$^{3}$Institute of Astronomy and Astrophysics, Academia Sinica, 11F of ASMAB, AS/NTU No.1, Sec. 4, Roosevelt Rd, Taipei 10617, Taiwan
}

\date{Accepted XXX. Received YYY; in original form ZZZ}

\pubyear{\the\year{}}

\begin{document}
\label{firstpage}
\pagerange{\pageref{firstpage}--\pageref{lastpage}}
\maketitle

\begin{abstract}
   We present the results of the further investigation of the Class~II methanol maser emission in the $14_1 - 14_0$~A$^{-+}$ transition {at 349.1~GHz} discovered in 2016 in the remarkable core S255IR-SMA1, harboring a $\sim$20~M\sun\ protostar NIRS3, which exhibited a disk-mediated accretion burst in 2015. 
   The present study is based on the observations of this object with ALMA in Band 7 at the largest baselines, which provide the angular resolution of $\sim$15~mas. We estimated physical conditions in the region from which comes the maser emission, and in the surroundings, using the presumably quasi-thermal methanol lines in our bands and the CH$_3$CN $19_\mathrm{K} - 18_\mathrm{K}$ line series.
   The total flux density in the $14_1 - 14_0$~A$^{-+}$ line in 2021 is about two times higher than in 2019. A maser emission of about the same intensity {in 2021} is {detected for the first time} in the $12_1 - 12_0$~A$^{-+}$ transition {at 336.9~GHz}. 
   The physical conditions in the masering and non-masering regions are similar. The masers 
   are apparently excited by the radiation of the central source. Unfortunately, the existing models cannot adequately take into account this radiation.
   The $18_{-3}-17_{-4}$~E transition at 345.919 GHz 
   shows characteristics of maser emission, too. 
   
\end{abstract}




\section{Introduction}

Molecular masers are an important diagnostic tool for star-forming regions. There are many known maser molecular transitions. There are also theoretical models, which explain excitation of these transitions and predict new ones. However, sometimes unpredicted masers are discovered. Such discoveries help to improve the molecular excitation models and provide new diagnostic capabilities. In 2016, we detected with ALMA a new, never predicted methanol maser line in the $14_1 - 14_0$~A$^{-+}$ transition at 349.1~GHz toward the well-known high-mass star-forming region S255IR \citep{Zin17}, soon after the disk-mediated accretion outburst in this object recorded in 2015 at NIR wavelengths, accompanied by the 6.7~GHz methanol maser flare and observed also in the radio domain \citep{Fujisawa15, Stecklum16, Caratti16, Moscadelli17, Szymczak2018, Cesaroni18}. Our ALMA observations in combination with the previous SMA results \citep{Zin15} indicate a submillimeter burst, which partly decayed in 2017 \citep{Liu18}. The methanol maser emission mentioned above, in 2017 decreased by 40\% in comparison with the 2016 level {($\sim$25~Jy)}, too \citep{Liu18}.

S255IR, at a distance of $1.78_{-0.11}^{+0.12}$~kpc \citep{Burns16}, is a well-known site of high-mass star formation \citep[e.g.,][]{Zinchenko2024vak}. The SMA1 core harbors a $\sim$20~M\sun\ protostar NIRS3 \citep{Zin15}, the mass of which is estimated from the bolometric luminosity of $\sim 3\times 10^4$~L\sun\ at the adopted distance.

In 2019, we observed S255IR in several lines of the CH$_3$OH $J_1 - J_0$~A$^{-+}$ series with the SMA \citep{Salii2022}. There was no clear evidence {in the line profile} for the maser component in the $14_1 - 14_0$~A$^{-+}$ line. The peak flux density was $\sim$5~Jy, which is about 5 times lower than in 2016 {and could be probably attributed to the thermal component}. At the same time, very likely, {based on the line profile analysis} we detected the maser component {($\sim$5~Jy)} in the $7_1 - 7_0$~A$^{-+}$ transition at 314.8~GHz. 

Also in 2019, we performed a survey of several high-mass star-forming regions in the lines of the CH$_3$OH $J_1 - J_0$~A$^{-+}$ series with the IRAM-30m radio telescope \citep{Salii2022}. In most cases there is no evidence of the maser emission except one component in NGC7538C, which may be a maser. Another line of this series in this source was suggested earlier to be a maser or an overheated transition \citep{Beuther2013}. The maser emission in the $14_1 - 14_0$~A$^{-+}$ line was definitely detected in G023.01--00.41 \citep{Sanna2021}.

\citet{Salii2022} theoretically analyzed conditions for maser excitation in the CH$_3$OH $J_1 - J_0$~A$^{-+}$ transitions. A simple model shows that the maser excitation can occur in quite hot clumps. 

In 2021, new ALMA observations of S255IR were carried out with a much higher angular resolution than earlier. In these observations the maser emission in the $14_1 - 14_0$~A$^{-+}$ line was detected, as well as in the $12_1 - 12_0$~A$^{-+}$ line {at 336.9~GHz} \citep{Zinchenko2024vak, Zinchenko2024}. Here we present and discuss the results of these observations.

\section{Observations}

The observations used in this paper are described in \citet{Zinchenko2024}. Briefly, we carried out our observations with the ALMA on 2021 September 3 toward S255IR SMA1 under the project \#2019.1.00315.S. The phase center in the ICRS reference frame was Right Ascension (RA) = 06$^\mathrm{h}$12$^\mathrm{h}$54{\fs}013 and Declination (Dec.) = $+$17\degr59\arcmin23{\farcs}050. The synthesized beam of the continuum map was $\sim 19\times 13$~mas, which corresponds to $\sim 34\times 23$~AU at the distance of S255IR. The synthesized beams for the spectral line maps are almost the same. The spectral resolution was 1.13~MHz ($\sim$1\kms).
Under the project \#2019.1.00315.S, we also carried out band 7 observations at $\sim$0{\farcs}1 resolution. Two executions were carried out on 2021 June 13 and July 6. The synthesized beam was $\sim 0.109\times 0.975$~arcsec.

\section{Results}

\begin{figure}
    \centering
    \includegraphics[width=0.9\linewidth]{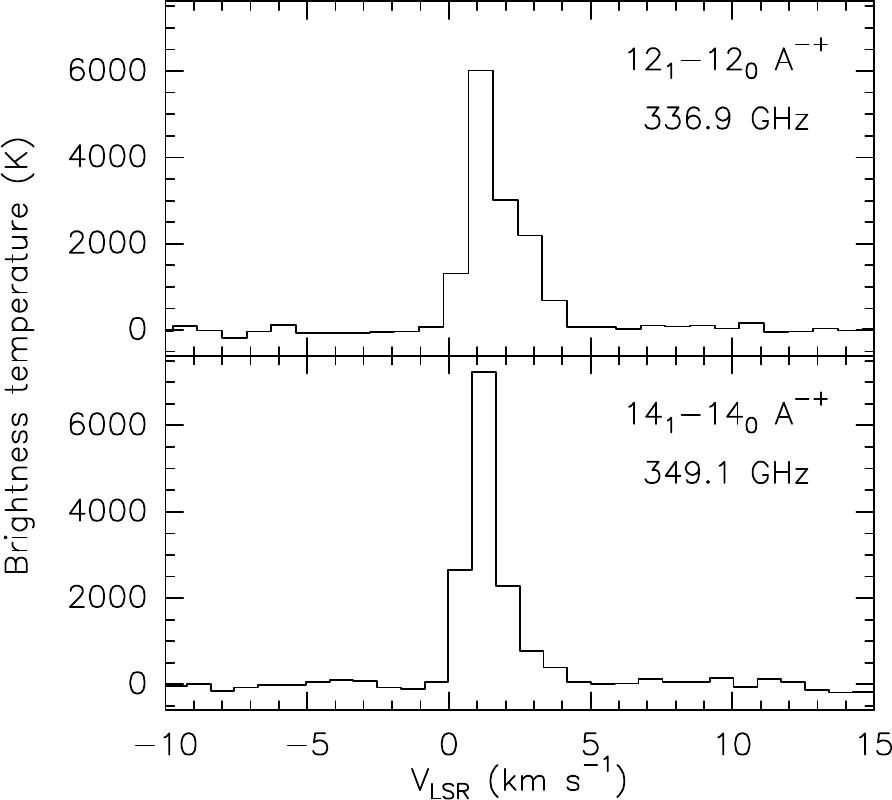}
    \caption{The spectra of the CH$_3$OH $12_1 - 12_0$~A$^{-+}$ (upper panel) and $14_1 - 14_0$~A$^{-+}$ (lower panel) emissions toward the brightness peak (R.A. = 06$^\mathrm{h}$12$^\mathrm{h}$54{\fs}0169, Dec. = $+$17\degr59\arcmin23{\farcs}101 {in the ICRS frame}).}
    \label{fig:spectra}
\end{figure}

\begin{figure}
    \centering
    \includegraphics[width=\linewidth]{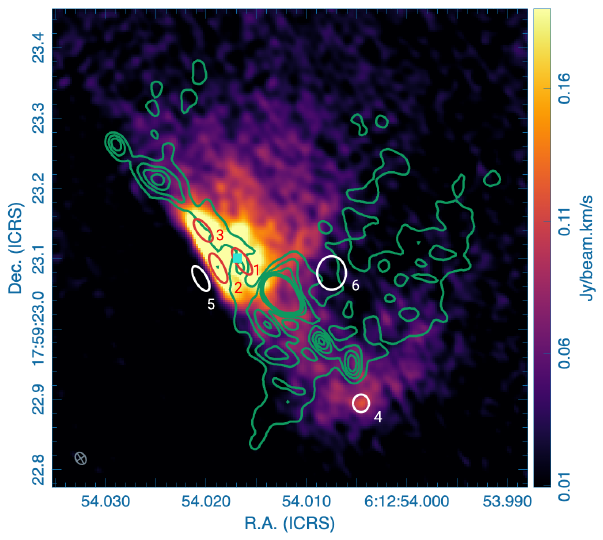}
    \caption{The map of the integrated intensity in the CH$_3$OH $12_1 - 12_0$~A$^{-+}$ line (image) overlaid with contours of the 0.9~mm continuum emission. The contour levels are 0.5 to 2.0 by 0.375~mJy\,beam$^{-1}$. {The numbered ellipses indicate the regions discussed in the text.} The cyan dot {inside Region 1} marks the brightness peak. 
    The synthesized beam is plotted in the lower left corner.}
    \label{fig:j12-map}
\end{figure}

\begin{figure}
    \centering
    \includegraphics[width=\linewidth]{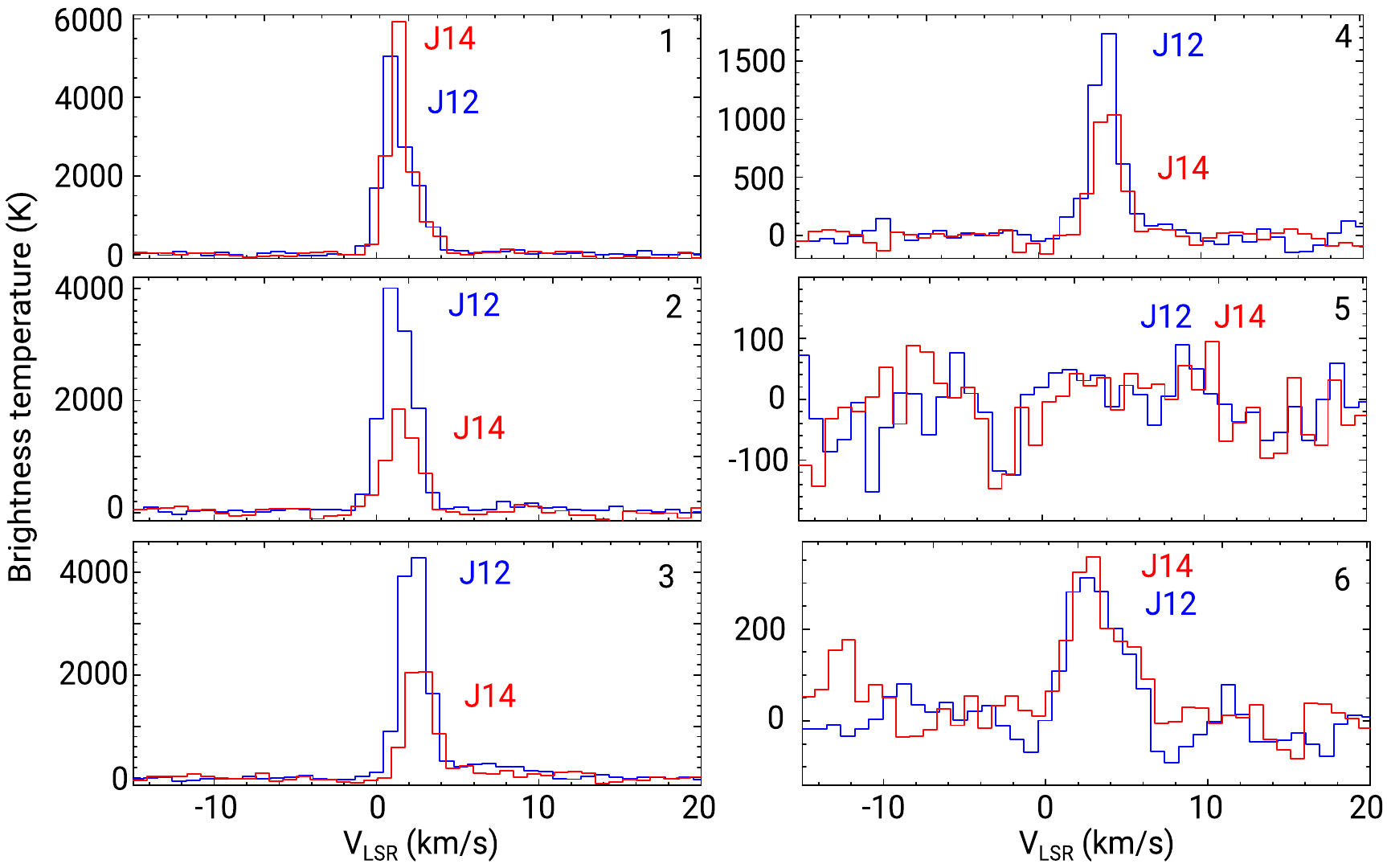}
    \caption{Spectra of the CH$_3$OH $12_1 - 12_0$~A$^{-+}$ and $14_1 - 14_0$~A$^{-+}$ lines in the regions indicated in Fig.~\ref{fig:j12-map}. The region number is shown in the upper left corner of each panel.}
    \label{fig:j12_regs}
\end{figure}

{The list of the observed methanol lines in our bands is presented in Table~\ref{tab:obs_methanol}.}
There are two lines of the CH$_3$OH $J_1 - J_0$~A$^{-+}$ series {among them}: $14_1 - 14_0$~A$^{-+}$ and $12_1 - 12_0$~A$^{-+}$. Both of them show a clear evidence of maser emission (Fig.~\ref{fig:spectra}). The brightness temperature reaches $\sim$7000~K. 
The map of the integrated intensity in the $12_1 - 12_0$~A$^{-+}$ line overlaid with contours of the 0.9~mm continuum emission is presented in Fig.~\ref{fig:j12-map}. 
{The spectra of both $12_1 - 12_0$~A$^{-+}$ and $14_1 - 14_0$~A$^{-+}$ transitions in the selected regions (Fig.~\ref{fig:j12-map}) are given in Fig.~\ref{fig:j12_regs}.}

\begin{table}
\centering
\caption{{Methanol transitions identified in our observations with a high angular resolution, according to CDMS \citep{2001A&A...370L..49M}}}
\label{tab:obs_methanol}
\begin{tabular}{cccc}
\hline
Frequency&  Transition     & $E_{u}$ &$\lg(A_E)$          \\
 MHz    &            & K & $\lg( \rm{s}^{-1})$\\                           
\hline
336438.224 	&$14_{7}-15_{6}$ A$^{++},\, v_t=0$ & 488.2  &$-4.44$\\
336605.889  &$7_{1}- 6_{1}$ A$^{++},\, v_t=2$  & 747.4  &$-3.79$\\
336865.149	&$12_{1}-12_{0}$ A$^{-+},\, v_t=0$ & 197.1  &$-3.39$\\
345903.916	&$16_{1}-15_{2}$ A$^{--},\, v_t=0$ & 332.6  &$-3.98$\\
345919.260 	&$18_{-3}-17_{-4}$ E, $v_t=0$    & 459.4  &$-4.13$\\
346202.719	&$5_{4}- 6_{3}$ A$^{--},\, v_t=0$  & 115.2  &$-4.66$\\
346204.271	&$5_{4}- 6_{3}$ A$^{++},\, v_t=0$  & 115.2  &$-4.66$\\
348031.826	&$7_{3}- 8_{2}$ E, $v_t=2$       & 686.2  &$-4.95$\\
349106.997	&$14_{1}-14_{0}$ A$^{-+},\, v_t=0$ & 260.2  &$-3.36$\\
\hline
\end{tabular}
\end{table}

\begin{figure}
    \centering
    \includegraphics[width=\linewidth]{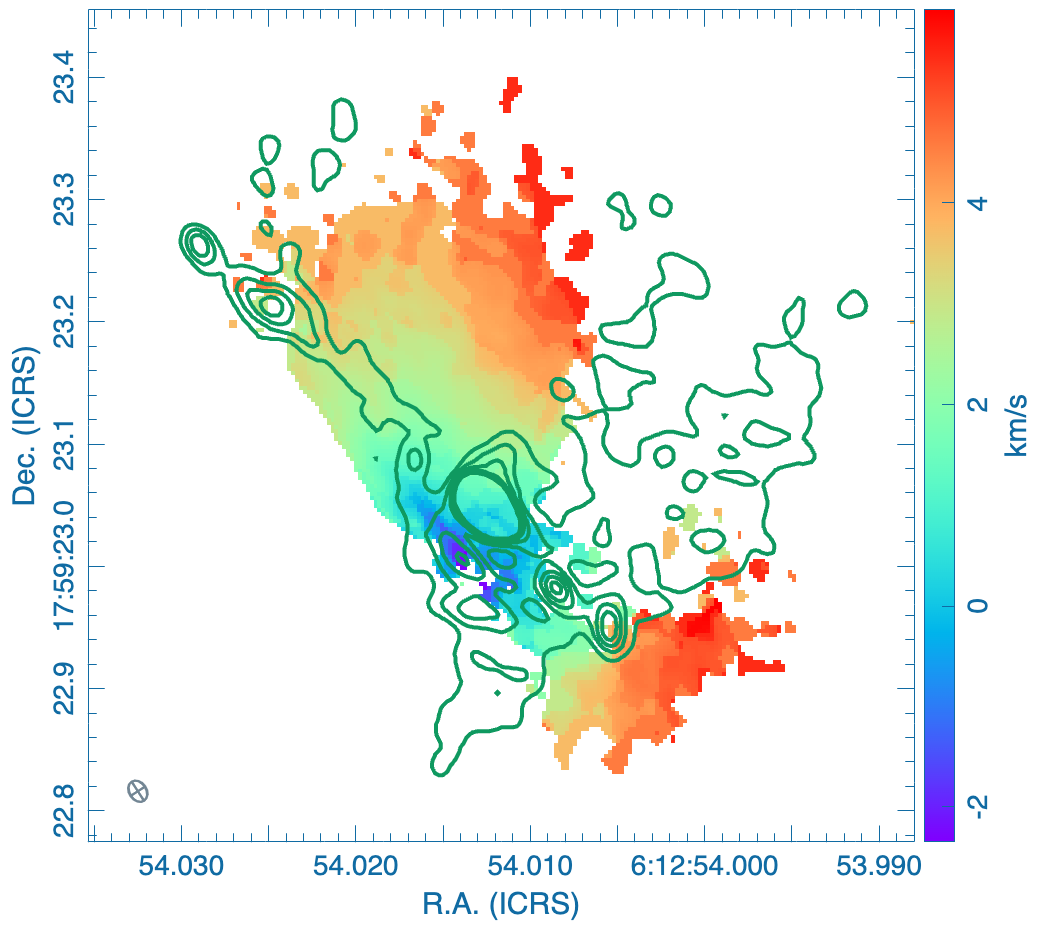}
    \caption{The {normalized} first moment map in the CH$_3$OH $12_1 - 12_0$~A$^{-+}$ line with the lower intensity cutoff of 15~mJy\,beam$^{-1}$, which corresponds to approximately 640~K. The contours are the same as in Fig.~\ref{fig:j12-map}.}
    \label{fig:j12-mom1}
\end{figure}

The strongest maser emission is observed along the NE (red-shifted) lobe of the jet described in \citet{Zinchenko2024}, in the area between the bright central continuum source and the NE2 knot in the jet. A weaker emission is observed also toward the SW (blue-shifted) lobe of the jet. There the emission peak is located ahead of the continuum knots.
There is a sharp drop of the maser emission to SE from the peak while in the NW direction the intensity decreases gradually. 
Kinematics of the masering region is illustrated by Fig.~\ref{fig:j12-mom1}, where the first moment map in the CH$_3$OH $12_1 - 12_0$~A$^{-+}$ line is plotted. The lower intensity cutoff is 15~mJy\,beam$^{-1}$, which corresponds to approximately 640~K. Intensities above this value most probably indicate a maser emission.

The peak flux density integrated over the whole area of the maser emission as seen at high resolution ($\sim$0{\farcs}60$\times$0{\farcs}35) is about 5.8~Jy in the $12_1 - 12_0$~A$^{-+}$ transition and about 5.0~Jy in the $14_1 - 14_0$~A$^{-+}$ transition (at $\sim$2.9\kms). Note that the brightness peak is observed at $\sim$1.4\kms\ (Fig.~\ref{fig:spectra}). In the low-resolution data the peak flux density in the $14_1 - 14_0$~A$^{-+}$ transition is about 10.0~Jy integrated in the region $\sim$1{\farcs}5$\times$1{\farcs}0. The $12_1 - 12_0$~A$^{-+}$ transition was not covered by the low-resolution observations.

The intensity ratio of the $12_1 - 12_0$~A$^{-+}$ and $14_1 - 14_0$~A$^{-+}$ lines varies across the emission region, as can be seen 
in Fig.~\ref{fig:j12_regs}. Around the brightness peak the intensities of both lines are almost equal, while near the SE boundary the $12_1 - 12_0$~A$^{-+}$ line is two times stronger than the $14_1 - 14_0$~A$^{-+}$ line.

\begin{figure*}
    \centering
    \includegraphics[width=0.48\linewidth]{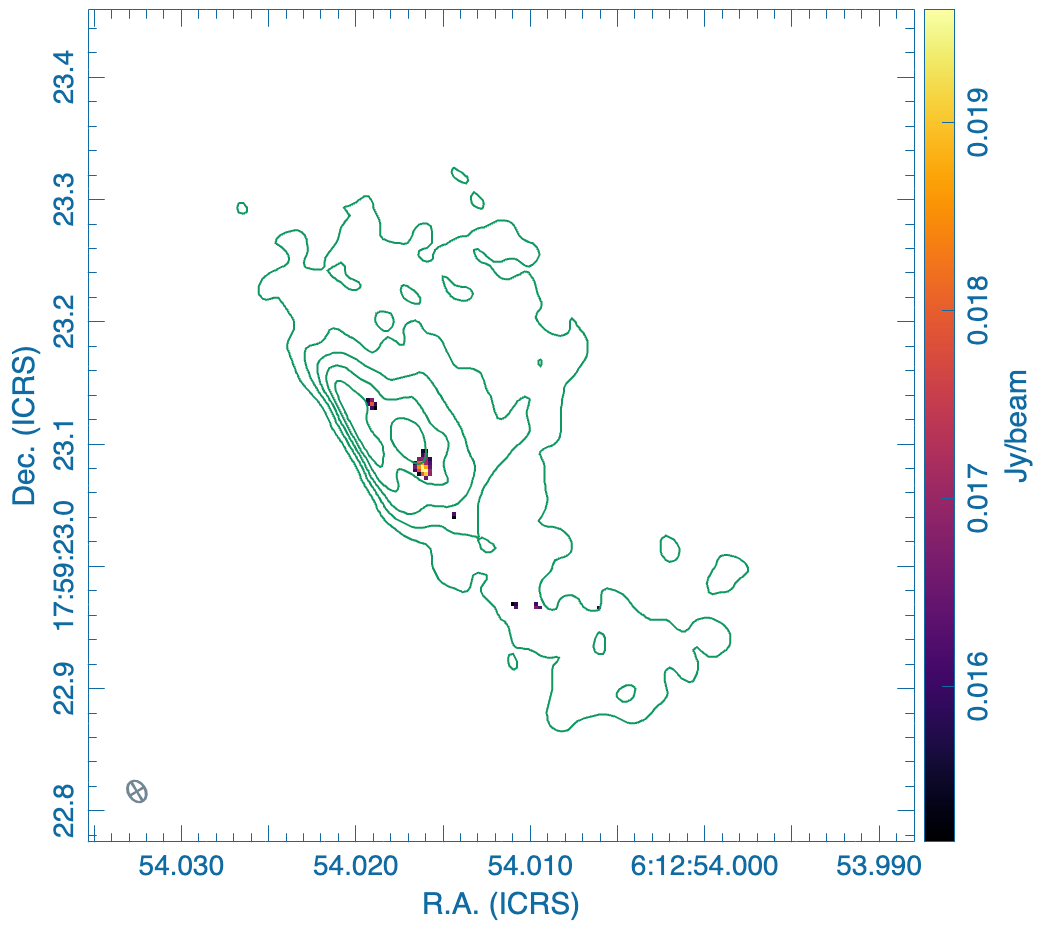}
    \hfill
    \includegraphics[width=0.48\linewidth]{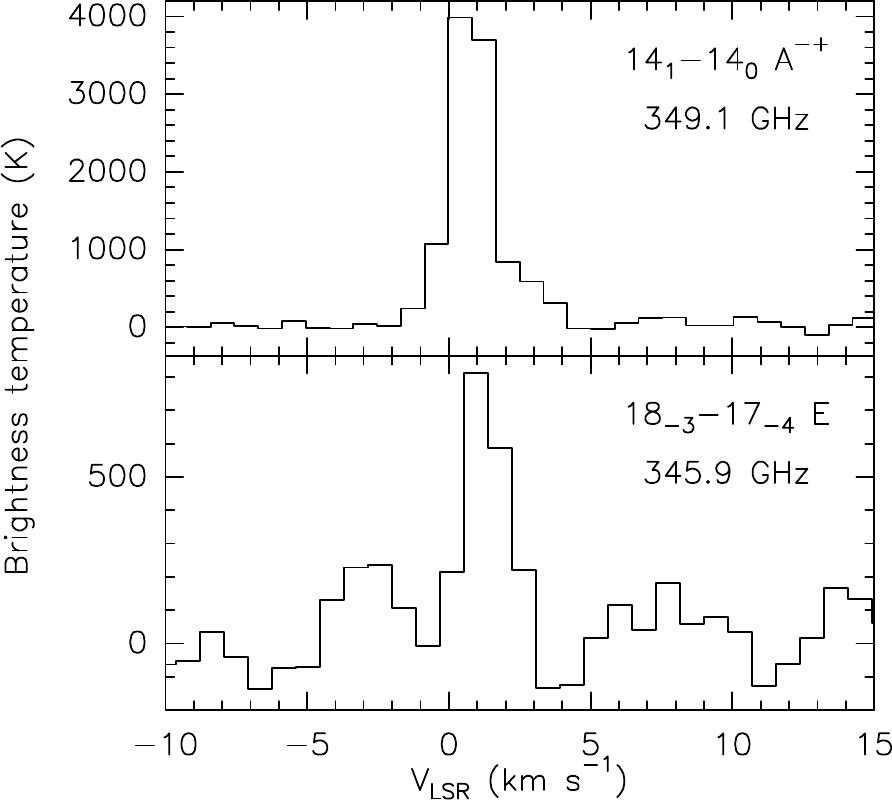}
    \caption{Left panel: the map of the peak intensity of the $18_{-3}-17_{-4}$~E line with the lower intensity cutoff of 15~mJy\,beam$^{-1}$, which corresponds to approximately 640~K. 
    The contours show the integrated intensity of the $12_1 - 12_0$~A$^{-+}$ maser line. The contour levels are from 0.05 to 0.25 by 0.05~Jy\,beam$^{-1}$\kms. Right panel: the spectra of the $18_{-3}-17_{-4}$~E line and $14_1 - 14_0$~A$^{-+}$ line toward the $18_{-3}-17_{-4}$~E emission peak.
    }
    \label{fig:345_maps}
\end{figure*}

We estimated physical conditions in the region of the maser emission and in the surroundings (Fig.~\ref{fig:j12_regs}) using the presumably quasi-thermal (hereafter thermal for simplicity) methanol lines in our bands and the CH$_3$CN $19_\mathrm{K} - 18_\mathrm{K}$ line series. We fitted the measured intensities of the 6 thermal methanol lines using the non-LTE approach described in \citet{Zin15}. However, some of these lines perhaps also show the maser effect. This is in particular the $18_{-3}-17_{-4}$~E transition at 345.919~GHz. The brightness in this line in the $J_1 - J_0$~A$^{-+}$ maser regions is significantly higher than predicted by our simple models 
(Fig.~\ref{fig:model}).
The spatial distribution of the peak brightness in this line (above $\sim$640~K) and its spectrum at the peak position are shown in Fig.~\ref{fig:345_maps}. The emission peak ($\sim$800~K) is located near the maser peak and the spectrum is similar to the $J_1 - J_0$~A$^{-+}$ maser line spectrum, demonstrating a narrow component with the width ($\sim$1.4\kms) considerably smaller than that of the thermal lines in this area ($\sim$3\kms). 
Therefore, it is very likely a maser, too.
The physical parameters derived from this modeling are rather similar for the regions with and without maser emission in the $J_1 - J_0$~A$^{-+}$ lines (Table~\ref{tab:mod_res_by_regs}). 

Estimates of the kinetic temperature from the CH$_3$CN $19_\mathrm{K} - 18_\mathrm{K}$ line series using the rotation diagram method {taking into account a high optical depth in the lines \citep[e.g.,][]{Liu2020}} are {similar to those obtained from the methanol lines ($\sim$250--400~K) and do not show a significant difference within the uncertainties between} the masering and non-masering regions. 

\section{Discussion}
The {total} flux density in the $14_1 - 14_0$~A$^{-+}$ line measured in 2021 with ALMA at high resolution is about the same as obtained in 2019 with the SMA and IRAM-30m \citep{Salii2022} but in the 2021 low-resolution data it is about two times higher. The line profiles integrated over the whole area of the maser emission are very similar at the two epochs. It shows that the emission in this line could increase on the time interval between 2019 and 2021 by a factor of $\sim$2. Of course, this emission is a sum of the maser and thermal components. It is hard to separate them. Most probably, the maser component was present in the 2019 data, too, although it could not be distinguished due to a low spatial resolution of those observations. It is worth noting that some spectral features of the 6.7~GHz maser emission show a gradual increase in this period, too, especially the new feature at 2.3\kms, which appeared in 2017 \citep{Aberfelds2023}.
In 2019, a rather strong maser component was evident in the $7_1 - 7_0$~A$^{-+}$ transition at 314.8~GHz \citep{Salii2022}. However, the results presented here show that other transitions of this series very likely could have a weaker maser component, too. 

The transverse velocity gradient in the maser lines across the jet may be related to the jet rotation. 
However, this gradient can be also attributed to rotation of this object as a whole. We do not discuss it here further.

\citet{Zin17} argued that the maser emission in the $14_1 - 14_0$~A$^{-+}$ line is a Class~II methanol maser. Recently one more Class II methanol maser ($6_1-5_2$ E at 265.3~GHz) was discovered in this area \citep{Baek2023}. The location, morphology and kinematics of this maser emission observed at about the same epoch are very similar to those shown in Figs.~\ref{fig:j12-map}, \ref{fig:j12-mom1}. Some other transitions observed in their work can also be Class~II masers. Our model described in \citet{Salii2022} shows that the $6_1-5_2$~E transition as well as the $18_{-3}-17_{-4}$~E transition mentioned above can be inverted at about the same physical conditions as the transitions of the $J_1 -J_0$ A$^{-+}$ series.

The similarity of the physical conditions in the regions with and without maser emission tells us that another factor should be important for the maser excitation. Moreover, the specific methanol column density derived from the thermal methanol lines is outside the range where the population inversion is possible according to the simple model considered in \citet{Salii2022}. In this model only the local physical conditions are taken into account. However, Figs.~\ref{fig:j12-map} and \ref{fig:j12-mom1} clearly show that the maser emission avoids the regions associated with a noticeable dust emission. Most probably, it means that the masers are excited by the radiation of the central source and are not observed in the regions which are shielded by dust. Unfortunately, the existing models cannot take into account this radiation because it has rather high colour temperature. This requires including highly excited levels and transitions which are not present in the currently available models of the methanol molecule structure. 

The peak of the maser emission is observed in a certain region of the NE jet lobe. However, the maser excitation is apparently not related to the jet, because in the SW lobe the maser peak is located ahead of the jet knots. The sharp drop of the maser emission to SE from the peak is probably explained by the presence of a cavity there created by previous ejections \citep{Zinchenko2024}. There is practically no other molecular emission in this area except the high-velocity CO. 

It is interesting that in the blue-shifted (SW) jet lobe we do not see a blue-shifted maser emission. Probably, this can be explained by a geometrical factor or the physical conditions in the blue-shifted gas, which prevent the maser excitation.

\section{Conclusions}

The main results of this study can be summarized as follows:

1. The peak brightness in the $14_1 - 14_0$~A$^{-+}$ methanol line in 2021 reached $\sim$7000~K at the $\sim$15~mas angular and $\sim$1\kms\ spectral resolutions. The peak flux density was about 10.0~Jy, which is 2 times higher than in 2019. A similar trend is observed in some components of the 6.7~GHz methanol maser emission. 

2. A maser emission of about the same intensity is detected in the $12_1 - 12_0$~A$^{-+}$ methanol transition. The intensity ratio of the $12_1 - 12_0$~A$^{-+}$ and $14_1 - 14_0$~A$^{-+}$ lines varies across the emission region by a factor of 2.

3. The physical conditions in the masering and non-masering regions 
are similar. The masers avoid the regions shielded by dust and are apparently excited by the radiation of the central source. Unfortunately, the existing models cannot adequately take into account this radiation.

4. The $18_{-3}-17_{-4}$~E transition at 345.919~GHz shows characteristics of maser emission, too (both narrow line width and high brightness temperature) toward several positions.

\section*{Acknowledgements}

{We are grateful to the anonymous referee for the useful comments.}
This work was supported by the Russian Science Foundation grants No. 24-12-00153 (https://rscf.ru/en/project/24-12-00153/; main text) and
No. 23-12-00258 (https://rscf.ru/project/23-12-00258/; Appendix).
This paper makes use of the following ALMA data: ADS/JAO.ALMA \#2019.1.00315.S. ALMA is a partnership of ESO (representing its member states), NSF (USA), and NINS (Japan), together with NRC (Canada), MoST and ASIAA (Taiwan), and KASI (Republic of Korea), in cooperation with the Republic of Chile. The Joint ALMA Observatory is operated by ESO, AUI/NRAO, and NAOJ.

\section*{Data Availability}

The {raw and pipeline processed} data are available in the ALMA archive {(ADS/JAO.ALMA \#2019.1.00315.S). The spectra and maps discussed in this paper are available by request from the corresponding author.}



\bibliographystyle{mnras}
\bibliography{s255ir-maser} 




\appendix



\section{Estimates of the physical parameters from thermal methanol lines} \label{sec:thermal}

We estimated physical parameters (gas kinetic temperature and density), methanol specific column densities, and abundances in several regions indicated in Fig.~\ref{fig:j12-map} on the basis of non-LTE modeling of several presumably thermal lines in the observed bands (Fig.~\ref{fig:j12_regs}). The modeling assumptions and methods are described in \citet{Zin15, Salii2022}.
It should be noted that our model uses a spherically symmetric LVG approximation 
and assumes equal gas and dust temperatures.  Whereas when modeling maser excitation, it is important to take into account the more complicated geometry of the source and a possible temperature difference between dust and gas within the source and its environment.

\begin{figure}
    \centering
    \includegraphics[width=\linewidth]{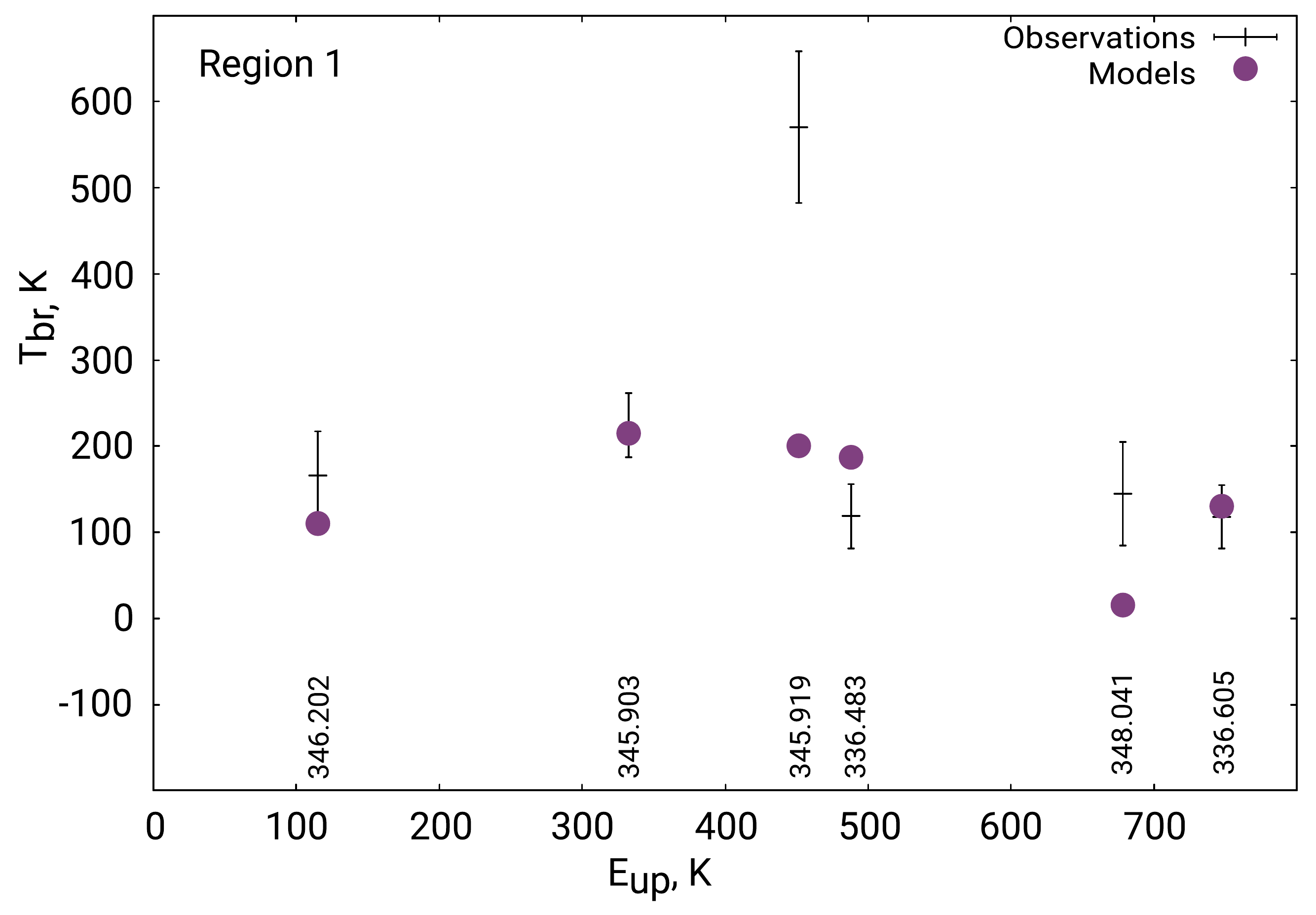}
    \includegraphics[width=\linewidth]{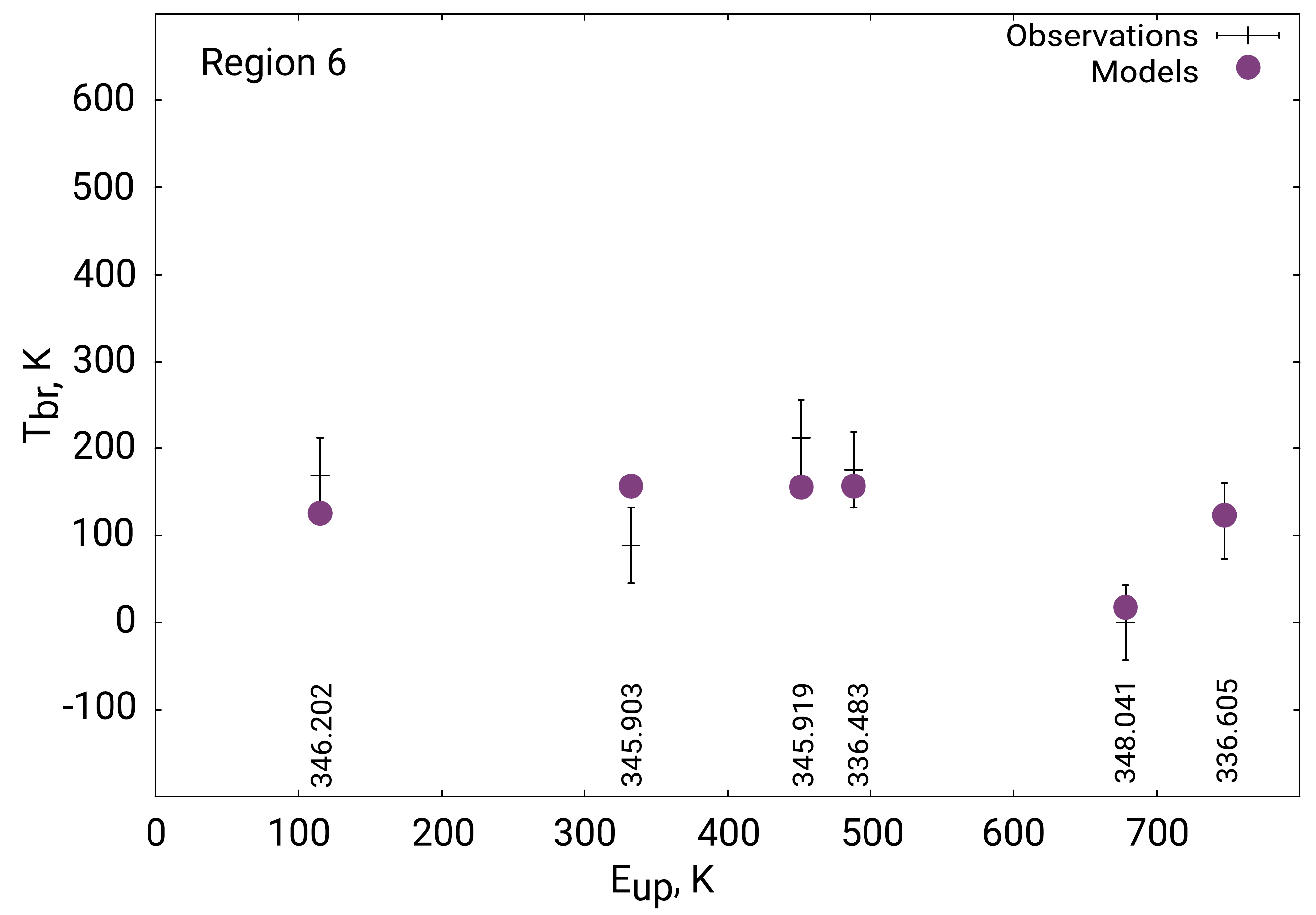}
    \caption{Examples of the modeling results in comparison with the measured line intensities for Regions~1 (upper panel) and 6 (lower panel). The  transition frequencies in GHz are shown near the bottom axis.}
    \label{fig:model}
\end{figure}

No methanol emission is detected in Region 5. Examples of the modeling results for two regions are presented in Fig.~\ref{fig:model}. Note that in Region~1 the measured intensity of the $18_{-3}-17_{-4}$~E transition is much higher than the model predictions. The same is true for the other regions with the detected methanol emission, except Region~6. 

The most likely values of the considered parameters and their 95\% confidence intervals are presented in Table~\ref{tab:mod_res_by_regs}. 

\begin{table*}
    \caption{The physical parameters (gas kinetic temperature and density), methanol specific column densities and abundances estimated from thermal methanol lines. The line width and filling factor were fixed at 1\kms\ and 100\%, respectively. The considered ranges of parameters are given in the second row of the header. The 95\% confidence intervals are presented in the parentheses.}
    \label{tab:mod_res_by_regs}
    \centering
    \begin{tabular}{cc c cc}
    \hline
 Region  & $T_\mathrm{k}$, K & $\lg(n_\mathrm{H_2}, \rm{cm^{-3}})$ & $\lg(N_\mathrm{CH_3OH}/\Delta V, \rm{s\,cm^{-3}})$ & $\lg(N_\mathrm{CH_3OH}/N_\mathrm{H_2})$ \\
        & 10\ldots600 & 3.00\ldots9.00       & 7.0\ldots14.0         & $-9\ldots-5.5$ \\
         \hline
 1 & 290(210\ldots500) & 8.25(4.00\ldots9.00) & 13.4(12.9\ldots14.0)  & $-5.5(-6.0\ldots-5.5)$ \\
 2 & 220(140\ldots400) & 9.00(4.00\ldots9.00) & 13.4(12.9\ldots14.0)  & $-5.5(-6.0\ldots-5.5)$ \\
 3 & 290(200\ldots280) & 5.00(4.00\ldots9.00) & 13.4(12.6\ldots14.0)  & $-5.5(-6.0\ldots-5.5)$ \\
 4 & 190(120\ldots410) & 5.00(4.00\ldots9.00) & 13.3(12.8\ldots14.0)  & $-5.5(-6.0\ldots-5.5)$ \\
 6 & 250(170\ldots420) & 5.00(4.00\ldots9.00) & 13.6(13.1\ldots14.0)  & $-5.5(-6.0\ldots-5.5)$ \\
 \hline
    \end{tabular}
\end{table*}

\bsp	
\label{lastpage}
\end{document}